\documentclass[aps,prr,showpacs,superscriptaddress,twocolumn,raggedfooter,raggedbottom,nofootinbib]{revtex4-2}

\usepackage{graphicx}
\usepackage{dcolumn}
\usepackage{bm}
\usepackage{amsmath,amssymb}
\usepackage{xcolor}

\def\p{\partial}

\usepackage{hyperref}

\begin{document}

\preprint{APS/123-QED}

\title{Quantum delay in the time of arrival of free-falling atoms}

\author{Mathieu Beau}
\affiliation{Physics department, University of Massachusetts, Boston, Massachusetts 02125, USA}
\affiliation{Massachusetts Institute of Technology, Cambridge, MA, USA}

\author{Lionel Martellini}

\affiliation{Finance department, EDHEC Business School, Nice 06200, France}

\date{\today}

\begin{abstract}
Using standard results from statistics, we show that for Gaussian quantum systems the distribution of a time measurement at a fixed position can be directly inferred from the distribution of a position measurement at a fixed time as given by the Born rule. In an application to a quantum particle of mass $m$ falling in a uniform gravitational field $g$, we use this approach to obtain an exact explicit expression for the probability density of the time-of-arrival (TOA). In the long time-of-flight approximation, we predict that the average positive relative shift with respect to the classical TOA in case of a zero initial mean velocity is asymptotically given by
$
\delta  = \frac{q^2}{2}
$
when the factor $q\equiv \frac{\hbar}{2m\sigma \sqrt{2gx}} \ll 1$ (semi-classical regime), and by
$
\delta  = \sqrt{\frac{2}{\pi}}q
$
when $q\gg 1$ (quantum regime), where $\sigma$ is the width of the initial Gaussian wavepacket and $x$ is the mean distance to the detector. We also discuss experimental conditions under which these predictions can be tested. 
\end{abstract}

\maketitle


While the Born rule gives the probability density of a position measurement at a fixed time, there is no readily available rule in the standard formalism of quantum mechanics for deriving the probability density of a time measurement at a fixed position. The origin of what is known as \textit{the first passage time problem} or \textit{the time of arrival (TOA) problem} in quantum mechanics can be traced down to the fact that there is no self-adjoint operator canonically conjugate to the Hamiltonian that can be associated with time measurements (\cite{pauli1933handbuch}).\footnote{The time of arrival operator in \cite{kijowski1974time,kijowski1999comment} is  self-adjoint but it does not strictly satisfy a canonical commutation relation with the Hamiltonian (see \cite{SCHLICHTINGER2023301}).}

The time of arrival problem holds particular importance in the context of free-falling particles, as an ongoing debate in the literature revolves around the existence of a potential deviation from the universality of free fall in the quantum domain. On the one hand, some authors invoke the correspondence principle to \textit{postulate} that the TOA should have a mean value agreeing with the classical value, and argue that it is only at the level of uncertainties that such a mass-dependence should occur (see for example \cite{Viola97,Davis04,Lämmerzahl96}). In contrast other authors argue in favor of a mass dependence also in the mean arrival time \cite{Flores19,Ali2006,Chowdhury2012}.

In this article we shed new light into the time of arrival problem and its application to free-fall. We follow an epistemic rather than an ontic approach (\cite{Bohr28} and \cite{Murdoch87}) in that we specifically focus on "measured time of arrival", that is the "external time" measured by a clock in the laboratory as per the terminology in \cite{Busch2008}, without touching upon the more fundamental question of the ontological nature of time in quantum mechanics. More precisely, we argue that a rule for obtaining the distribution of a time measurement at a fixed position is actually readily available within the standard formalism. For Gaussian systems, we show that it can in fact be directly inferred from the distribution of the position measurement at a fixed time as given by the Born rule. 
To see this, we first define $X_{t}$ as the random variable associated with the \textit{measured} position at a fixed time $t$, and symmetrically define $T_{x}$ as the random variable associated with the \textit{measured} time of passage at a fixed position $x$. Using standard results from statistics, we are then able to obtain the probability distribution of $T_{x}$ as a function of the probability distribution of $X_{t}$ (see equation \eqref{Eq:TOA:Distribution:Gaussian}). In an application to the free-falling quantum particle, we find that (i) the mean time of arrival is greater than the classical time of arrival as the result of a Jensen's inequality, and that (ii) it is a function of the mass of the particle, among other ingredients. We also provide an analytical expression for the asymptotic positive shift of the mean TOA with respect to the classical particle in both the semi-classical and quantum regimes (see equations \eqref{Eq:delta}-\eqref{Eq:deltaquantum}). These predictions can be empirically tested in experimental conditions that we discuss below.  

Many attempts have already been made in the literature to address this fundamental question (see \cite{muga2007time,Ruschhaupt15} for reviews of previous work) but a consensus is yet to be reached. In a nutshell, two main bodies of research have analyzed the TOA problem within the standard formulation of quantum mechanics, while a third stream of research has explored the question within the context of Bohmian mechanics. The first approach consists in \textit{randomizing} the classical equation of motion via the introduction of uncertain initial position and velocity conditions as per the position/momentum uncertainty relationship. This pragmatic (sometimes also referred to as ``hybrid'') approach is particularly well-suited for the analysis of experimental results since it can accommodate the presence of various physical constraints in the practical implementation of the measurement process (see \cite{dufour2014shaping} or \cite{Rousselle22} for recent examples of analyses of time of arrival of free-falling anti-hydrogen atoms in the context of the GBAR experiment at CERN). One conceptual limit of this approach, however, is that it is not truly quantum mechanical and ignores the full information embedded in the dynamical propagation of the state of the system through the Schrödinger equation. One of the contributions of this article is in fact to show that for a free-falling particle the randomization of the classical equation of motion can be regarded as a long time-of-flight approximation to the exact solution (see equation \eqref{Eq:approx:TOA} and the discussion that follows). The second approach consists in \textit{quantizing} the classical equation of motion in an attempt to obtain a quantum time-of-arrival operator. Beside the intrinsically \textit{ad hoc} nature of the search for a quantum time operator, this approach does not easily lead to explicit results regarding the TOA distribution. For example, the authors of \cite{Rovelli96} only deal with the free particle, for which they are simply able to obtain a semi-classical approximation. The free fall problem is analyzed via a dedicated TOA operator in \cite{Flores19}, but the authors are also unable to obtain analytical solutions for the TOA distribution or even its moments, and the associated eigenvalue problem is solved numerically by coarse graining. Another contribution of this article is to provide in contrast a general exact analytical expression for the TOA distribution that is valid for all Gaussian quantum systems (see equation \eqref{Eq:TOA:Distribution:Gaussian}). In the application to free fall, and specializing to the long-time-of-flight situation, we are also able to obtain an analytical asymptotic expressions for the mean and the standard-deviation of the TOA in both the semi-classical and quantum regimes. We use these results, which to the best of our knowledge are new to the literature, to estimate the relative shift with respect to the classical time-of-arrival (see equations \eqref{Eq:delta} and \eqref{Eq:deltaquantum}). Finally the third approach to the TOA problem involves a departure from the standard and accepted formalism since it consists in casting the problem within Bohmian mechanics. Interestingly enough, the explicit expression we obtain for the TOA distribution of the free-falling particle (see equation \eqref{Eq:TOA:Distribution:Freemotion}) coincides with the result obtained with Bohmian mechanics (see \cite{leavens1998time} and \cite{ali2006quantum}). Another contribution of this article is to provide a more general expression that applies to any Gaussian state (see equation \eqref{Eq:TOA:Distribution:Gaussian}) while staying within standard formalism of quantum mechanics, and also to discuss the conditions of validity for this expression and how to proceed when these conditions are not satisfied.\\


\textit{Stochastic measured position at a fixed time ($X_t$).}  In quantum mechanics, the one-dimensional time-dependent Schrödinger equation sets the dynamics of the wavefunction $\Psi_t(x)$:
\begin{equation}\label{Eq:Schro}
    -\frac{\hbar^2}{2m}\frac{\p^2}{\p x^2}\Psi_t(x) +V(x,t)\Psi_t(x) = i\hbar \frac{\p}{\p t}\Psi_t(x), \
\end{equation}
where $m$ is the mass of the particle and $V(x,t)$ is a position- and time-dependent external potential. By the Born rule, the density of probability for the particle to be measured in a small region around the position $x$ at a given time $t$ is given by: 
\begin{equation}\label{Eq:rho}
   \rho_t(x) \equiv  |\Psi_t(x)|^2.\ 
\end{equation}
As indicated earlier, we denote by $X_t$ the random variable associated to the Gaussian probability density function $\rho_t(x)$. In other words $X_t$ is defined to represent the uncertain outcome of a first measurement performed at date $t$ after the system has been prepared in the state represented by $\Psi_0$ and has evolved, according to the Schrödinger equation and with no prior measurement, to the state $\Psi_t$. 

From a principle standpoint, this position measurement corresponds to the following experiment: (i) we place $n$ detectors at different positions $x_1,\ x_2,\ \cdots,\ x_{n}$ with a spatial resolution $\delta x$ in an interval $[a,b]$ (hence, $x_0=a,\ x_1=a+\delta x,\ \cdots, x_k = a+k\delta x,\ \cdots,\ x_{n}=b$), (ii) we synchronize these detectors to make sure they turn on at the same exact time $t$, (iii) and then we record the position of the particle at the time $t$. From equation \eqref{Eq:rho}, the probability that the k-th detector detects the particle at the time $t$ is $P_k(t) = \int_{x_k}^{x_{k+1}}|\Psi_t(x)|^2 dx$. (iv) We repeat this procedure a large number $N$ of times, which allows us to reconstruct the density of probability $\rho_t(x)$ at a fixed time $t$. At each trial $p=1,\ 2,\ 3,\cdots,\ N$, we measure the value of the position $x$ of the particle at a given time $t$, and we can thus represent this outcome as the realization of a stochastic variable $X_t$ that gives the measured position $x\in[a,b]$ of the particle at the time $t$, for which the density distribution is given by $\rho_t(x)$. 

While our approach is more general, we specialize in what follows the analysis to Gaussian states with a density distribution of the form
\begin{equation}\label{Eq:density:gaussian}
    \rho_t(x) = \frac{1}{\sqrt{2\pi \sigma(t)^2}}e^{-\frac{(x-x_c(t))^2}{2\sigma(t)^2}},\
\end{equation}
where $\sigma(t)$ is the standard-deviation of the Gaussian distribution that is centered at the classical path $x_c(t)$, which by the correspondence principle is also the mean value of the position operator $\langle \hat{x}_t\rangle = \int_{-\infty}^{+\infty}x\rho_t(x) = x_c(t)$). Gaussian states are standard forms in quantum physics, not only for the free fall problem which is the focus of this article, but also for the free motion, the simple and time-dependent harmonic oscillator, constant or time-dependent electric fields, and more generally for any quadratic potential of the form $V(x,t) = a(t)x^2+b(t)x$, where $a(t)$ and $b(t)$ are two functions of $t$ (see for example \cite{Klebert73}). 
We further denote the initial mean value of the position by $x_0$ and its initial standard-deviation by $\sigma$. In the Gaussian setting, $X_t$ can be written with no loss of generality as:
\begin{equation}\label{Eq:rv:gaussian:t>0}
    X_t  = x_c(t) + \xi\sigma(t),\ 
\end{equation}
where $\xi=\mathcal{N}(0,1)$ is a normally distributed random variable with a variance of $1$ and a mean value of $0$.\\

\textit{Stochastic measured time at a fixed position ($T_x$)}. In the previous thought experiment, we considered that the $n$ position-detectors were turned on at a fixed time $t$. This means that the detectors are synchronized to an ideal clock that allows them to switch on through a signal at a precise time-value $t$. 
We now turn to a symmetric perspective, where the focus is on time measurements at a fixed position. 
As recalled in the introduction, there exists an abundant literature on the subject that has generated a number of insightful results that sometimes contradict each other because of implicit divergences in the underlying definition of a time measurement. In this context, we seek to avoid ambiguities by carefully explaining the experimental setup that would be involved in an idealized yet physical measurement of the TOA. Specifically, we consider the following procedure: (i) we place a single detector at the fixed position $x$; (ii) we drop a particle at time $0$ and we turn on the detector (say by triggering a laser pulse) at some time $t$; (iii) we record $1$ if the particle has been measured at position $x$ for this particular time $t$ and $0$ otherwise. Then, (iv) we repeat the steps (i)-(iii) $N$ times while keeping the exact same time $t$ and we count the total number of particles detected at this position (alternatively, we could in principle use in step (ii) an atomic cloud with $N$ non-interacting particles \cite{Dalibard95,Dalibard96}). Finally, (v) we repeat the steps (i)-(iv) by letting $t$ vary, with a small enough temporal resolution $\delta t$ (hence, $t_0=0,\ t_1=\delta t,\ \cdots, t_k = k\delta t,\ \cdots,\ t_n=n\delta t$). This procedure allows us to reconstruct the whole time distribution $\Pi_x(t)$ of a random variable, denoted by $T_x$ (note the symmetry in notation with respect to $X_t$), which can be regarded as a stochastic time of arrival (STOA) at the fixed position $x$. Please note that this approach differs from a procedure that would consist in performing continuous measurements (with a detector placed at position $x$) starting at $t=0$ and until the first detection is recorded. Indeed, such a procedure would involve multiple measurements before the detection occurs, in contradiction with our definition of $T_x$, which is set to be the random date of a \textit{first} measurement at position $x$.

The question that naturally arises at this stage is to find the expression of the STOA as a function of the position $x$ of the detector. Formally, $T_x$ is defined as the first passage time of the random process $X_t$ at the position $x$:
\begin{equation}\label{Eq:defTOA}
    T_x \equiv \text{inf} \left\{ t | X_t=x\right\} 
\end{equation}
As explained above, we can experimentally determine the possible values of $T_x$ by fixing the position of the detector at $x$ and allowing the time of observation to vary. From \eqref{Eq:defTOA} these values correspond to the solution of the equation $x = X_{T_x}$, where $X_t$ is given in \eqref{Eq:rv:gaussian:t>0}. Hence, we obtain the following mapping between the random variable $\xi$ and the STOA $T_x$ to be:
\begin{equation}\label{Eq:rv:map:Tx}
    x  = x_c(T_x) + \xi\sigma(T_x),\
\end{equation}
This equation can be rewritten as 
\begin{equation}\label{Eq:map:h}
    T_x = h_x(\xi),\
\end{equation}
assuming the existence of an invertible function where $h_x(\cdot)$ such that 
\begin{equation}\label{Eq:TOA:xi:General}
\xi = h_x^{-1}(T_x) = \frac{x-x_c(T_x)}{\sigma(T_x)},\
\end{equation}
 (see \eqref{Eq:rv:map:X0T} for an approximate expression of the function $h_x$ for the free-falling particle).
Assuming that the function $h_x$ is strictly monotonic, a standard result from statistics, sometimes referred to as the \textit{method of transformations}, gives the following relation between the probability distribution $\Pi_x(t)$ for the STOA $T_x$ at the detection point $x$ and the probability distribution $f(\cdot)$ of the standardized Gaussian variable $\xi$ (see for example theorem 4.1 in Chapter 4.1.3 in \cite{StochTextbook}):
\begin{equation}\label{Eq:TOA:Distribution:General}
    \Pi_x(t) = 
    f(h_x^{-1}(t))\times \left|\frac{\p}{\p t}h_x^{-1}(t)\right|. \
\end{equation}
Note that this result can in fact be extended to a more general case by relaxing the assumption of strict monotonicity. Indeed, if $h_x$ is not monotonic, one can usually partition its domain of definition into a finite number of intervals such that it is strictly monotonic and differentiable on each partition.
 Finally, we use
$$
\frac{\p}{\p t}\left(\frac{x-x_c(t)}{\sigma(t)}\right)\ = -\left(\frac{v_c(t)\sigma(t)+(x-x_c(t))\dot{\sigma}(t)}{\sigma(t)^2}\right),
$$
where $v_c(t)$ is the classical velocity and $\dot{\sigma}(t) = d\sigma(t)/dt$, we find the following expression for the time-of-arrival distribution:
\begin{equation}\label{Eq:TOA:Distribution:Gaussian}
    \Pi_x(t) =  \left|\frac{v_c(t)\sigma(t)+(x-x_c(t))\dot{\sigma}(t)}{\sigma(t)^2}\right|\times \frac{1}{\sqrt{2\pi}}e^{-\frac{(x-x_c(t))^2}{2\sigma(t)^2}}.\
\end{equation}
To the best of our knowledge, this general expression for the density distribution of the TOA for a Gaussian system is new to the literature. 
In what follows we discuss the application to a free-falling particle, and also to the free particle as a nested case with a zero gravitational potential. \\


\textit{Time-distribution of a free-falling particle.} For the free-falling particle, we recall the standard expressions for the classical path  $x_c(t) = v_0 t + \frac{g}{2}t^2$  (fixing the mean initial position $x_0=0$ and assuming $g>0$) and for the standard-deviation $ \sigma(t) = \sigma \sqrt{1+\frac{t^2}{\tau^2}}$, where $\tau = \frac{2m\sigma^2}{\hbar}$ is a characteristic time. Here $v_0>0$ represents the mean value \footnote{We assume $v_0$ to be positive so that the function $h_x$ in equation \eqref{Eq:map:h} is strictly monotonic, which implies that equation \eqref{Eq:TOA:Distribution:General} is valid. For the case where $v_0<0$, we will need to partition the domain of definition of $h_x$ in order to write the correct expression of equation \eqref{Eq:TOA:Distribution:General}. This goes beyond the scope of this paper and will be examined in a follow up paper.} for the initial velocity of the particle, which itself is a random variable with a standard-deviation given from the uncertainty principle as $\sigma_v=\hbar/2m\sigma$.  Using \eqref{Eq:TOA:Distribution:Gaussian} we obtain the exact expression for the probability distribution of the STOA for the free-falling particle:
\begin{equation}\label{Eq:TOA:Distribution:Freemotion}
    \Pi_x(t)
    = \left(\frac{v_0 +x\frac{t}{\tau^2}+gt+\frac{g}{2}\frac{t^3}{\tau^2}}{1+\frac{t^2}{\tau^2}}\right)\times \frac{\exp{\left(-\frac{(x-v_0 t-\frac{gt^2}{2})^2}{2\sigma^2\left(1+\frac{t^2}{\tau^2}\right)}\right)}}{\sqrt{2\pi\sigma^2} \sqrt{1+\frac{t^2}{\tau^2}}}, \
\end{equation}
where the distribution is normalized $\int_{-\infty}^{+\infty} \Pi_x(t)dt =1$. 
Here we have assumed that the function $h$ exists so that the solution $T_x$ to equation \eqref{Eq:map:h} takes on positive values with probability 1, but this assumption may not always hold. For example, if $\xi>x/\sigma$, the particle would start moving forward from an initial position already greater than $x$, implying that there is no time $t>0$ such that it would cross the position $x$ again. However we can restrict our attention to experimental situations where $x\gg\sigma$, and hence focus on cases where there is always a positive value of $t$ solution to equation \eqref{Eq:map:h}. In such situations we can assume that the probability for the particle to reach $x$ for $t<0$ is negligible and we can therefore consider that $\int_{0}^{+\infty} \Pi_x(t)dt \approx\int_{-\infty}^{+\infty} \Pi_x(t)dt=1$. Notice that when $v_0=0$ this formula coincides with the one obtained in \cite{ali2006quantum} from the quantum current density approach derived from the Bohmian approach (see also \cite{leavens1998time}). Our result is more general in that we derive the distribution for $v_0 \geq 0 $ and find the explicit expression for the normalization factor. Also, as mentioned before, we could use our approach to (i) explore the near-field regime ($x\sim \sigma$) and (ii) investigate the case where the function $h_x$ in equation \eqref{Eq:map:h} is not monotonic (e.g., if $v_0<0$). 

From equation \eqref{Eq:TOA:Distribution:Freemotion}, we remark that all the $n$-th moments $\int_{-\infty}^{+\infty} t^n\Pi_x(t)dt <+\infty$ converge.  There is no analytical expression for such integrals, but they can easily be computed numerically. In what follows, we use an alternative approach consisting in first finding an approximate explicit expression for the stochastic time-of-variable $T_x$ as a solution to equation \eqref{Eq:rv:map:Tx}, and then obtaining analytical estimates for its mean and standard deviation. The solution to equation \eqref{Eq:rv:map:Tx} for the free-fall:
\begin{equation}\label{Eq:FF}
    x = v_0 t + \frac{g}{2}t^2 + \sigma \xi \sqrt{1+\frac{t^2}{\tau^2}}, \ 
\end{equation}
is indeed difficult to express in a closed form but we can find an approximation for this expression in the semi-classical regime for long time-of-flight $t\gg \tau$.  \\

\textit{Time-of-arrival of a free-falling particle in the long time-of-flight regime} 
In the long time-of-flight regime where $t\gg \tau$, equation \eqref{Eq:rv:map:Tx} for the free-falling particle becomes:
\begin{equation}\label{Eq:approx:TOA}
    x \approx \frac{gT_x^2}{2} + v_0 T_x + \sigma\xi\frac{T_x}{\tau}, \ 
\end{equation}
which is exactly equal to the classical trajectory the particle would have at time $T_x$ with an initial position equal to $0$ and an initial velocity equal to $v_0 + \frac{\sigma\xi}{\tau}$. 
The use of the classical equation with uncertain initial conditions, which is at the core of the ``hybrid'' approach, can thus be regarded for the free-falling particle as an approximation valid in the long time-of-flight case. In the short time of flight, and more generally for other Gaussian quantum systems, we should instead revert back to the exact equation \eqref{Eq:TOA:Distribution:Gaussian} in order to calculate the distribution and moments of the TOA.  

The solution to equation \eqref{Eq:approx:TOA} is given by a standard quadratic formula, from which we find the stochastic time-of-arrival variable to be:
\begin{equation}\label{Eq:rv:map:X0T}
    T_x = h_x(\xi) = \frac{1}{g}\left(\sqrt{\left(v_0+\frac{\sigma\xi}{\tau}\right)^2+2gx}-\left(v_0+\frac{\sigma\xi}{\tau}\right)\right).\
\end{equation}
Notice that for $\sigma=0$, we find the classical time:
$
    t_c = \frac{1}{g}\left(\sqrt{v_0^2+2gx}-v_0\right).\
$    
Consider now the semi-classical condition $q\ll 1$, where the factor $q$ measures the ratio of the height-dependent wavelength to the initial width of the particle wave-packet
\begin{equation}\label{Eq:q}
    q = \frac{\sigma}{\sqrt{v_0^2+2gx}\ \tau} = \frac{\hbar}{2m\sigma\sqrt{v_0^2+2gx}} = \frac{\lambda}{2\pi\sigma},\
\end{equation}
and where $\lambda = h/(m\sqrt{v_0^2+2gx})$. In this semi-classical case, we obtain
\begin{equation}\label{Eq:TOA:longTOF:sc:freefall}
    T_x\approx t_c - \frac{t_c}{\sqrt{v_0^2+2gx}}\frac{\sigma\xi}{\tau}+\frac{x\sigma^2\xi^2}{\left(v_0^2+2gx\right)^{3/2}\tau^2}.\
\end{equation}
We can see from equation \eqref{Eq:TOA:longTOF:sc:freefall} that $T_x$ is a strictly convex function of $\xi$, which implies by the Jensen's inequality that its mean value $\mathbb{E}(T_x)$ is strictly greater than the classical time $t_c$ as $\mathbb{E}(T_x) = \mathbb{E}(h_x(\xi))>h_x(\mathbb{E}(\xi))=h_x(0)=t_c$.\footnote{The Jensen's inequality is a standard tool in statistics, widely used in various fields of application. Simply put, it states that the expected value of a strictly convex function of a random variable is strictly greater than 
the convex function of the expected value of the random variable.}
Moreover it is easy to obtain from equation \eqref{Eq:TOA:longTOF:sc:freefall} the following expressions for the mean value and standard deviation of the TOA (see Appendix \ref{Appendix} for further details):
\begin{equation}\label{Eq:TOA:longTOF:sc:meanvar:freefall}
    \begin{cases}
        t_{\text{mean}} \approx t_c + \frac{x\sigma^2}{(v_0^2+2gx)^{3/2}\tau^2}\\
        \\
        \Delta T_x  \approx t_c\frac{\sigma}{\sqrt{v_0^2+2gx}\tau}
    \end{cases}.\
\end{equation}

\begin{table*}[t]
\begin{ruledtabular}
  \begin{tabular}{ccccccc}
    \hline
    Atoms  &  $x$ ($\text{m}$) & $g$ ($\text{m}\cdot\text{s}^{-2}$) & $t_c$ ($\text{s}$)   & $\delta t $ ($\text{s}$) & $q$ & $\delta$  \\ \hline
    Hydrogen-1 & $0.1$  &  $9.81$ & $0.143$ & $0.004$ & $ 0.225 $ & $ 2.5\% $  \\ 
    Hydrogen-1 & $0.01$  &  $9.81$ & $0.045$ & $0.009$ & $ 0.713 $ &  $ 20\% $  \\ 
    Hydrogen-1 & $0.1$  &  $10^{-5}$ & $141$ & $2.5053\times 10^4$ & $ 223 $ & $ 1.77\times 10^4\% $ \\ 
  \end{tabular}
  \caption{\textbf{Experimental prediction of quantum delay for the free fall of atoms}. In this table, we show the expected time-shift $\delta t$ for a free-falling particle using typical values for the hydrogen-1 atom in the ground state of a harmonic trap with an angular frequency of $\omega = 3.16$MHz (i.e, the initial width of the wave packet is $\sigma = 10^{-7}$m) in three different situations: (i) on earth with the standard experimental conditions, (ii) on earth with a shorter distance traveled $x$, and (iii) in micro-gravity with the same $x$ as in (i). }\label{Fig:Table}
\end{ruledtabular}
\end{table*}

It is worth mentioning that the expressions of the mean value and the standard deviation of the TOA for the free motion can be obtained by taking the limit $g\rightarrow 0$ (which implies that $t_c\rightarrow x/v_0$) in the expressions \eqref{Eq:TOA:longTOF:sc:meanvar:freefall}. 
Notice that the expression for the time-distribution \eqref{Eq:TOA:Distribution:Freemotion} in this limit ($g\rightarrow 0$) is the same as equations (A9) and (A18) in \cite{Rovelli96}. The difference here is that we obtain the formula as an exact expression, while the authors of \cite{Rovelli96} present this result as an approximation to the order of $\frac{\sigma}{v_0\tau}$ in the semi-classical regime.
Other authors have also found a similar formula in the semi-classical regime using a quantum flux approach (see for example equation (9) in \cite{Muga98}). These results are consistent with the finding reported in \cite{Flores19} that the expectation value of the TOA operator for a free-falling particle is equal to the classical time of arrival plus mass dependent quantum correction terms.

Interestingly, if the particle is dropped in the gravity field ($v_0=0$), we find that the mean value of the fall is greater than the classical value $t_c = \sqrt{\frac{2x}{g}}$ by a relative factor:
\begin{equation}\label{Eq:delta}
    \delta = \frac{t_{\text{fall}}-t_c}{t_c} \approx 
        \frac{q^2}{2} = \frac{\hbar^2}{16gxm^2\sigma^2}\ . 
\end{equation}

Turning to the quantum regime ($q\gg 1$), we can rewrite the TOA in equation \eqref{Eq:rv:map:X0T} as: 
$$
T_x = qt_c \left(\sqrt{\xi^2+\frac{1}{q^2}}-\xi\right) \approx  qt_c \left(|\xi|-\xi\right)
$$
when $q\gg1$. Hence, we obtain in the full quantum regime the following estimate for the relative deviation with respect to the classical TOA:
\begin{equation}\label{Eq:deltaquantum}
   \delta = \sqrt{\frac{2}{\pi}}q = \sqrt{\frac{2}{\pi}}\frac{\hbar}{2\sqrt{2 gx}m\sigma}\ ,
\end{equation}
where we used $\int_{-\infty}^{+\infty}d\xi\ |\xi|\frac{e^{-\xi^2/2}}{\sqrt{2\pi}} = \sqrt{\frac{2}{\pi}}$. To the best of our knowledge, this expression has never been derived in the literature.

\begin{figure}
    \centering
    \includegraphics[scale=0.5]{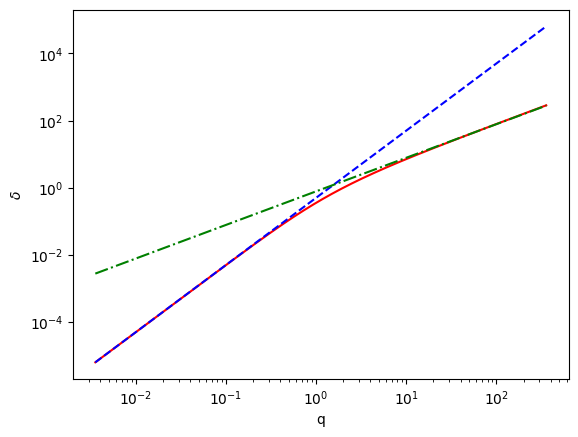}
    \caption{\textbf{Relative quantum delay versus quantumness.} In this figure, we show the value of the relative quantum delay defined by $\delta = (t_{\text{fall}}-t_c)/t_c$ as a function of the coefficient of quantumness $q=\frac{\hbar}{2m\sigma\sqrt{2gx}}$ in the long time-of-flight (TOF) regime (the TOF varies from $700 \tau$ to $2 \times 10^{10} \tau$) in the log-log scale.  
    For comparison, we display the asymptotic values obtained from the numerical integration of the mean of the distribution \eqref{Eq:TOA:Distribution:Freemotion}  (continuous-red-line), as well as the values obtained in the semi-classical regime as per equation \eqref{Eq:delta}  (dashed-blue-line), and within the quantum regime as per equation \eqref{Eq:deltaquantum} (dashed-dotted-green-line).} 
    \label{fig:delta}
\end{figure}

These results are surprising for at least two reasons: (i) the mean time of free fall is not equal to (is strictly greater than) the classical time, and (ii) it depends on the mass of the particle. In Figure \ref{fig:delta}, we show that the expressions \eqref{Eq:delta} and \eqref{Eq:deltaquantum} give excellent approximations for the numerical values of the integral $\int_{0}^{+\infty}t\Pi_x(t)dt$ (which we calculate using the scipy.integrate library of python 3.10) when the coefficient of quantumness $q$ is much lower or much greater than $1$, respectively. In the transition towards the quantum regime, that is when $q\sim1$, the analytical approximation deteriorates with respect to the more accurate numerical integration. However, our explicit expressions \eqref{Eq:delta}-\eqref{Eq:deltaquantum} provide a convenient tool to help search for optimal experimental conditions to observe this quantum delay in the semi-classical and quantum regime. In particular we propose in Table \ref{Fig:Table} three different scenarios for a realization of the quantum-delay experiment. The first line of Table \ref{Fig:Table}  shows that the effect is small ($2.5\%$) but noticeable for a sufficiently sensitive detector. Fortunately, one can follow two different strategies to make the deviation significant enough to be more easily detectable: (i) either we decrease the distance traveled (second line of Table \ref{Fig:Table}) or (ii) we realize a microgravity experiment (third line of the table). Some technical challenges must be addressed before one can reach these experimental conditions, but we believe that the potential benefits of measuring a deviation from the universality of free fall make the investigation worth pursuing.

\textit{Discussion.} 
Using standard results from statistics, we introduce a novel approach to analyze the time-of-arrival for a free-falling particle. We derive the probability density for the stochastic time-of-arrival (STOA) $T_x$ and obtain analytical expressions for its mean and its standard derivation in the semiclassical regime and in the quantum regime for a long-time flight. In further research, we might extend our analysis to encompass the regime of short-time of flight, where $t \ll \tau$, as well as the near-field scenarios where $x \sim \sigma$. Our approach can also be extended in a straightforward manner to any other Gaussian system, which is a setting general enough to encompass a number of important standard systems including the free motion, the free-fall, and the harmonic oscillator (simple and time-dependent).
While this present article only considers applications to Gaussian states, our equations \eqref{Eq:TOA:xi:General} and \eqref{Eq:TOA:Distribution:General} actually provide a general framework that can be applied to the study of entangled particles \cite{Lloyd01,Lloyd02}, quantum superposition \cite{Roncallo23}, quantum gases \cite{Gaaloul22}, 
potential barriers \cite{Dumont93,Leavens93}, two-slit experiment \cite{Frabboni12}, diffraction in time \cite{Moshinsky52,Dalibard96,delCampo09}, and quantum backflow \cite{Allock69,Bracken94,YearsleyHalliwell13}.

We find that the mean TOA for a free-falling object is not equal to the classical time. Although the time shift can be very small, we propose various experimental scenarios where this subtle effect could be measured. Specifically, we consider situations involving atoms with small mass, short distances traveled (these two conditions could be met in the GBAR experiment\cite{Rousselle22}), and microgravity. A future space mission (see e.g., STE-QUEST \cite{Altschul15}) with low-mass atoms emerges as a highly promising candidate for investigating and unraveling this phenomenon.

\textbf{Acknowledgments.} We are grateful to Prof. Jacob Barandes, Dr. Simone Colombo, and Prof. Seth Lloyd for helpful comments and fruitful discussions. This research was conducted while Lionel Martellini was a visiting professor at MIT.

\newpage

\bibliography{PRAsubmv3.1}

\newpage
\onecolumngrid

\begin{center}
\textbf{\large Appendix}
\end{center}

\section{Derivation of equations (17) and (18)}\label{Appendix}

In the long time-of-flight regime where $t\gg \tau$, we have:
$$
    x \approx \frac{gt^2}{2} + v_0 t + \sigma\xi\frac{t}{\tau}, \
$$
which is exactly equal to the classical trajectory the particle would have with an initial position equal to $0$ and an initial velocity equal to $v_0 + \frac{\sigma\xi}{\tau}$. 
The solution to this equation is given by a quadratic formula, thus we find the stochastic time-of-arrival variable to be:
\begin{equation}\label{SM:Eq:TOA:longTOF:freefall}
    T = \frac{1}{g}\left(\sqrt{\left(v_0+\frac{\sigma\xi}{\tau}\right)^2+2gx}-\left(v_0+\frac{\sigma\xi}{\tau}\right)\right).\ 
\end{equation}
Notice that for $\sigma=0$, we find the classical time:
$$
    t_c = \frac{1}{g}\left(\sqrt{v_0^2+2gx}-v_0\right).\
$$

Consider now the semi-classical regime $\sigma \ll v_0 \tau$. 
Since
$$
\left(v_0+\frac{\sigma\xi}{\tau}\right)^2+2gx = v_0^2+2gx+\frac{2v_0\sigma\xi}{\tau}+\frac{\sigma^2\xi^2}{\tau^2},\
$$
and given that the first two terms $v_0^2+2gx$ are very large compared to the next two ones, we find the Taylor series:
\begin{align*}
    \sqrt{\left(v_0+\frac{\sigma\xi}{\tau}\right)^2+2gx} &\approx \sqrt{v_0^2+2gx}\left(1+\frac{v_0\sigma\xi}{(v_0^2+2gx)\tau} +\frac{\sigma^2\xi^2}{2(v_0^2+2gx)\tau^2}-\frac{v_0^2\sigma^2\xi^2}{2(v_0^2+2gx)^2\tau^2}\right)\\
    & = \sqrt{v_0^2+2gx}\left(1+\frac{v_0\sigma\xi}{(v_0^2+2gx)\tau} +\frac{gx\sigma^2\xi^2}{2(v_0^2+2gx)^2\tau^2}\right) ,
\end{align*}
where we used $\sqrt{1+a}\approx 1+\frac{a}{2}-\frac{a^2}{8}$, when $a\ll 1$.
We can now approximate \eqref{SM:Eq:TOA:longTOF:freefall} as:
\begin{equation}\label{SM:Eq:T:approx}
T\approx \frac{1}{g}\left(\sqrt{v_0^2+2gx}-v_0\right) + \left(\frac{v_0}{\sqrt{v_0^2+2gx}}-1\right)\frac{\sigma\xi}{g\tau}+\frac{x\sigma^2\xi^2}{\left(v_0^2+2gx\right)^{3/2}\tau^2}.\
\end{equation}
Notice that the classical time is given by:
\begin{equation}\label{SM:Eq:tc:freefall}
    t_c = \frac{1}{g}\left(\sqrt{v_0^2+2gx}-v_0\right),\
\end{equation}
as expected. Thus, we can rewrite the previous equation \eqref{SM:Eq:T:approx} as:
\begin{equation}\label{vEq:TOA:longTOF:sc:freefall}
    T\approx t_c - \frac{t_c}{\sqrt{v_0^2+2gx}}\frac{\sigma\xi}{\tau}+\frac{x\sigma^2\xi^2}{\left(v_0^2+2gx\right)^{3/2}\tau^2}.\
\end{equation}
The second order approximation of the squared value of $T$ is given by:
$$
T^2 \approx t_c^2 + \frac{t_c^2}{v_0^2+2gx}\frac{\sigma^2\xi^2}{\tau^2} + \frac{2t_c x\sigma^2\xi^2}{(v_0^2+2gx)^{3/2}\tau^2} - \frac{2t_c^2}{\sqrt{v_0^2+2gx}}\frac{\sigma\xi}{\tau}.\
$$
Using the last two expressions, we find the formulae for the mean and the variance of $T$:
\begin{equation}\label{SM:Eq:TOA:longTOF:sc:meanvar:freefall}
    \begin{cases}
        t_{\text{mean}} = E(T) \approx t_c + \frac{x\sigma^2}{(v_0^2+2gx)^{3/2}\tau^2}\\
        \\
        V(T) = E(T^2) - E(T)^2 \approx t_c^2\frac{\sigma^2}{(v_0^2+2gx)\tau^2}
    \end{cases},\
\end{equation}
whence the expression of the standard deviation of $T$:
\begin{equation}\label{SM:Eq:TOA:longTOF:sc:STD:freefall}
    \Delta T \approx t_c\frac{\sigma}{\tau\sqrt{v_0^2+2gx}}.\
\end{equation}

Notice that when $g\rightarrow 0$, we obtain the standard-deviation for the TOA of the free particle. Interestingly, if the particle is dropped in the gravity field ($v_0=0$), we have:
$$
t_{\text{mean}} \approx \sqrt{\frac{2x}{g}} + \frac{1}{2}\sqrt{\frac{2x}{g}}\frac{\sigma^2}{2gx\tau^2}= \sqrt{\frac{2x}{g}}\left(1+\frac{\sigma^2}{4gx\tau^2}\right) =\sqrt{\frac{2x}{g}}\left(1+\frac{\hbar^2}{16gxm^2\sigma^2}\right)
$$
leading to equation (18).


\end{document}